# B-Call: Integrating Ideological Position and Political Cohesion in Legislative Voting Models

Juan Reutter[1], Sergio Toro[2], Lucas Valenzuela Everke[3], Daniel Alcatruz[4] and Macarena Valenzuela[5]

**Abstract**

This paper combines two significant areas of political science research: measuring individual ideological position and cohesion. Although both approaches help analyze legislative behaviors, no unified model currently integrates these dimensions. To fill this gap, the paper proposes a methodology called B-Call that combines ideological positioning with voting cohesion, treating votes as random variables. The model is empirically validated using roll-call data from the United States, Brazil, and Chile legislatures, which represent diverse legislative dynamics. The analysis aims to capture the complexities of voting and legislative behaviors, resulting in a two-dimensional indicator. This study addresses gaps in current legislative voting models, particularly in contexts with limited party control.

**Keywords**

Ideological Position, Coherence, B-Call, W-Nominate

[1] Departamento de Ciencia de la Computación. Pontificia Universidad Católica de Chile

[2] Escuela de Gobierno y Administración Pública, Universidad Mayor, Chile

[3] Instituto Milenio de Fundamento de los Datos

[4] Instituto Milenio de Fundamento de los Datos

[5] Departamento de Sociología y Ciencia Política, Universidad Católica de Temuco, Chile

**Corresponding Author:**

Sergio Toro Maureira. Universidad Mayor, Chile

Correo: sergio.toro@umayor.cl.

## 1. Introduction

Legislative behavior measurement models have been widely used in political science since the second half of the 20th century. This development marks at least two predominant lines of research. The first analyzes ideological positioning based on voting, using tools that place parliamentarians on a left-right continuum within Euclidean space (Poole and Rosenthal 1985). The second approach identifies party cohesion and/or dicipline. This type of study is especially fruitful in creating aggregated indicators that assess the ability of political groups to maintain control in decision-making (Rice, 1925; Carey, 2000, 2007).

Both tools have been efficient in characterizing the various legislative systems around the world, especially those that are more stable and have greater party control. However, these techniques have also encountered measurement issues, particularly when dealing with ambivalent individual behaviors and low cohesion in voting (Carrizosa, 2023). This article addresses this complexity by proposing a two-dimensional tool that manages to integrate each legislator's ideological positioning with their level of cohesion concerning their peers. This model has been named B-Call.

B-Call, that is an abbrevation of Bidimensional Analisys of Roll Call, assumes that votes are random variables in the interval $[-1,1]$, where $-1$ represents rejection of a vote, $0$ abstention and $1$ approval. Consider a legislator $i$, and let $v_{ij}$ be the vote of legislator $i$ in the voting $j$, and $M_j$ is the average of all votes for the voting $j$. Then the average of the values $v_{ij} - M_j$ over all the voting $j = 1,...$, is normally distributed, associating a different curve to each legislator. Consequently, we argue both theoretically and empirically that the *average* of the Gaussian curve of each legislator is related to their ideological position, while the standard deviation is a good estimator of a legislator's level of cohesion. This means that a low *standard deviation* represents a legislator who votes according to their ideological position (cohesive behavior), whereas a high standard deviation indicates that the legislator's votes are less cohesive with their ideological position (volatile bahavior).

To test the B-Call model, nominal voting data from the congresses of the United States, Brazil, and Chile are used. These cases were chosen for their academic tradition in measuring legislative behavior and positioning, as well as their similarities and differences in political cycles. Indeed, the three countries have experienced both periods of stability and fragmentation. On the other hand, they exhibit different legislative dynamics. The U.S. Congress, for example, is characterized by two dominant political parties. Brazil, on the other hand, shows a high level of fragmentation in its two legislative chambers. Meanwhile, Chile is transitioning from a system dominated by two coalitions to one that is strongly fragmented. These similarities and differences allow testing the efficacy of B-Call in different contexts.

The article is divided into four parts. The first is a theoretical approach to legislative behavior measurement models, discussing their application in ideological dimensions and voting cohesion. The second part formalizes the measurement model to integrate analysis in both dimensions. The third part involves testing the B-Call model against others traditionally used to analyze legislative behavior. The fourth part is empirical and aims to highlight the advantages of B-Call in terms of adaptability in different political contexts. Finally, a discussion on the scope of this methodology and its future uses is developed.

## 2. Literature Review

Although there are several more, the measurement of legislative behavior has two main approaches: a) ideological positioning, b) Intra-Party Cohesion (IPC). In these lines of research, measurement models based on nominal roll-call votes in the full congress have been presented. For example, regarding ideological positioning, the models correspond to the spatial theory of voting and the individual utility functions of legislators. On the other hand, the measurement of cohesion results from some type of aggregation of preferences within Congress.

The approaches related to ideological positioning are based on voting patterns, particularly in spatial voting models. These measurements argue that the ideological position of legislators can be inferred through the study of the belief system summarized in an ideological range (Poole & Rosenthal, 1985, 1987, 1997; Londregan, 2000; Jackman, 2001; Clinton, Jackman & Rivers 2004; Quinn, 2004). There are two widely recognized models that rely on a utility function, assigning choice probabilities to each vote according to the alternatives available in each roll-call vote. These models, known as "NOMINATE" or "IDEAL," aim to assess a legislator's decision-making within a set of alternatives that maximize their utility. Both models assume that the legislator assigns a utility to voting for one option over another. In particular, the utility associated with each alternative is determined in part by the spatial distance between the vote and the position most preferred by a legislator; the closer the alternative is to the preferred position, the higher the utility assigned by the legislator. The random utility shocks in both models can be distributed as an extreme value, representing unobserved factors that could affect the choice.

In the case of NOMINATE proposed by Poole & Rosenthal (1985), multidimensional analysis techniques are used to position legislators and votes within a Euclidean political space. The NOMINATE model is based on an agnostic approach to estimate the ideal points of legislators from their voting patterns, treating each vote as a latent expression of their preferences in one or more dimensions. W-NOMINATE introduces improvements by considering the relative importance of different votes, allowing for their differential weighting in the analysis. Similarly, DW-NOMINATE is an evolution of the original W-NOMINATE, designed to track voting patterns over time, incorporating a dynamic component (Poole & Rosenthal, 1997). In the case of W-NOMINATE, the utility function has a Gaussian form intended to capture the probability that a legislator will vote for a particular bill. On the other hand, the IDEAL indicator is an approximation to the NOMINATE utility function with the implementation of Bayesian methods, which, unlike methods with latent positions, proposes to observe how preferences have been distributed in votes to estimate the relationship between the ideal ideological position for a voting pattern. Studies by Quinn (2004) and the works together with Clinton, Jackman, & Rivers (2004) represent Bayesian approaches in the analysis of voting behaviors that have been very useful for positioning the preferences of congress members.

This line of research has significantly contributed to improving the understanding of decision-making dynamics in Congress. However, it still faces challenges due to its limited capacity to capture the interaction between individual decisions and the complexities of partisan dynamics. Bateman & Lapinski (2016) point out that the agnosticism of these models generates estimates that do not allow for adequately observing the development of policies and contexts. Similarly, Krehbiel & Peskowitz (2015), demonstrate that the type of legislative organization is a condition that causes biases in the estimation of these indicators. Zucco Jr & Lauderdale (2011) find something similar in Brazil, suggesting the need for additional information to distinguish the influence of executive behavior and the setup of roll-call votes in specific policies in changing coalitions. Caughey & Schickler (2016) also highlight the limitations of the indicators to analyze conflicts and ideological changes over time, proposing the use of specific subsets of votes according to the historical context and the interest of analysis.

The second approach in the analysis of legislative behavior focuses on Cohesion. This approach has enabled the detection of the level of party unity and the relative importance of discipline in the effectiveness of decisions (Krehbiel, 2000). One of the most well-known indicators in this field is the RICE index (Rice, 1925). Subsequently, other indicators were developed that have been well regarded for their versatility and adaptability in different legislatures and political contexts. Carey (2000, 2007), for example, analyzed the "UNITY", "RICE", and "WELDON" indicators in 10 presidential countries, verifying their usefulness for reaching empirical conclusions regarding the proportion of seats and the levels of party discipline. Similarly, more recent studies (Ramirez, 2009; Carson et al., 2010; Stecker, 2015; Wang & Peng, 2015; Itzkovitch-Malka & Hazan 2017; Dingler & Ramstetter, 2023; Mai & Wenzelburger, 2023; Moskowitz & Rogowski 2024) examine the relationship between party unity and phenomena such as electoral outcomes, internal party conflicts, the emergence of new legislative agendas, the gender effect, the importance of legislative design, and institutional changes such as electoral rule changes.

Indicators of party discipline and cohesion are essential for assessing political systems in terms of stability and functionality. However, these tools have limitations, as they are susceptible to the nature of the set of votes they aim to use or measure (Crespin, Rohde, & Wielen, 2013). Therefore, it is possible that the scores of party unity may be influenced by changes in the legislative agenda or by events accompanying the votes. This is particularly complex in contexts where levels of aggregation are volatile or where the party system is not entirely stable.

Although there are efforts to integrate both dimensions, these efforts still seem isolated. For example, the approach studied by Aldrich, Rohde, & Tofias (2007) on Conditional Party Government (CPG) complements this line of work by better evidencing some nuances of the historical periods of partisan strength and division in Congress. From ideal point estimation using DW-NOMINATE between 1877-2002 they note that some components such as inter-party heterogeneity, intra-party homogeneity, separation, and party label adequacies can be derived. They combined these measures using factor analysis to derive a single CPG score for each congress. A two-dimensional strategy of ideal scores would reveal more information about partisan structuring in relation to the historical big occurrences of congressional preferences.

On the other hand, Bianco & Sened (2005) seeks to complement this perspective by examining the influence of the majority party in Congress through the characterization of the uncovered set in legislative proceedings. This analysis allows modeling legislative decision making in a much more realistic way, providing information on the potential agenda-setting and influence of the majority party. The emphasis of this study is on the importance of a multidimensional analysis to better identify some assumptions about preference biases from strategic agenda setting or polarization.

Thus, given the difficulties of measurement in changing and fragmented contexts, new techniques are required for the study of political dynamics. While traditional spatial voting analysis techniques provide valuable insights into the ideological dimensions of political systems, they also have limitations in understanding the multiplicity of ideological and strategic nuances within Congress. This last point is what we aim to address using the B-Call model, which integrates, in two dimensions, ideological positioning with voting cohesion. The construction of this model is presented below.

## 3. B-Call: A Two-Dimensional Model

B-Call treats votes as variables that take values $-1, 0, 1$, where $-1$ indicates the rejection of a vote, 0 abstention, y 1 approval. We use $v_{ij}$ to represent the vote of legislator $i$ in vote $j$, where $i \in I$ and $j \in J$, with $I$ being a set of legislators y $J$ a set of votes. Likewise, we use $M_j \in [-1,1]$ to represent the average of the votes of each legislator in vote $j$.

Our model is based on studying, for each legislator $i$, the sequence of values $(v_{ij} - M_j)$, which represents how much legislator $i$ deviates from the general framework of a particular vote $j$. However, for this value to also be an estimator of the legislator's ideology, a bit more work is needed. Initially, legislators are divided into two groups, theoretically labeled as *left* and *right*. Then, a vote $j$ is considered *right* when $M_{lj} < M_{rj}$: the average $M_{lj}$ of the votes from the left-wing group is less than the average $M_{rj}$ from the right-wing group (i.e., the left group tended to reject that vote more than the right group). Conversely, the vote is *left* when $M_{lj} \geq M_{rj}$. For each vote, a function $j$ is defined as $f_j: [-1,1] \to [-1,1]$ as $f_j(x) = x$ if $M_{lj} < M_{rj}$ (the vote $j$ is right-wing) or $f_j(x) = -x$ if $M_{lj} \geq M_{rj}$ (the vote is left-wing). The function $f_j$ is used as a transformation to determine if the legislator leaned more to the left or right than the average: if $f_j(v_{ij} - M_j)$ is negative, it means one of the following two statements is true: either $v_{i,j} < M_j$ and $M_{lj} \leq M_{rj}$, meaning that legislator $i$ rejects a vote where the left tended to reject more than the right, or $v_{i,j} \geq M_j$ and $M_{lj} \geq M_{rj}$, meaning that legislator $i$ approves a vote in which the left tended to approve more than the right. Intuitively, in both cases, the legislator voted more towards the left than average. Likewise, a positive value $f_j(v_{ij} - M_j)$ indicates that the legislator tended to approve a vote in which the right tended to approve as much or more than the left, or that legislator $i$ tended to reject a vote where the right rejected as much or more than the left. In other words, the legislator voted more towards the right than average.

Our model, then, assumes that the votes $v_{i1}, v_{i2}, \ldots, v_{ij}, \ldots$ of legislator $i$ are random events (according to some unknown distribution), and moreover, that the values $f_j(v_{ij} - M_j)$ for the same legislator $i$ are independent and distributed according to the same random distribution. Thus, the central limit theorem ensures that the distribution of these magnitudes converges to a normal distribution, upon which we can estimate its mean and variance. More specifically, and adding a standardization factor, the dimensions of the model consist of estimating the mean and standard deviation of the indicator $u_{ij}$ for legislator $i$ and vote $j$, which results from the standardization of the vote $v_{ij}$ with the average of the vote $M_j$ and the standard deviation $S_j$:

$$u_{ij} = \begin{cases} \dfrac{v_{ij} - M_j}{S_j} & if\ M_{lj} \leq M_{rj} \\ -\dfrac{v_{ij} - M_j}{S_j} & otherwise \end{cases}$$

Based on these assumptions, the sequence of values $u_{i1}, \dots, u_{iJ}$ provides a good estimate for the mean and variance of the underlying Gaussian distribution for each legislator. These two measures correspond to the two dimensions of B-Call, and can be used to estimate the ideological position of legislators, as well as the cohesion of their votes.

### 3.1. First Dimension: Ideological Position

The first dimension of the model corresponds to the average of the $u_{ij}$ values for each legislator $i$. As mentioned, this average is an estimator of the mean of the normal distribution associated with this legislator. This dimension is associated with legislator $i$"s ideological position in the political center. A negative or positive sign of $d_{1i}$ places legislator $i$ on the left or the right, respectively, and the magnitude of $d_{1i}$ indicates how far this legislator is from the political center.

$$d_{1i} = \frac{1}{n} \sum_{j=1}^{n} u_{ij}$$

### 3.2. Second Dimension: Cohesion

Given that the $u_{ij}$ values converge to a normal distribution, it is relevant to consider the standard deviation. For example, consider figure1. This figure shows the probability density function of two different Gaussian curves, estimated from the actual votes of two deputies elected by the same party in the Chilean congress. On the $x$-axis, we locate the ideological position of each vote, estimated as the average of the $u_{ij}$, values, and on the $y$-axis, the probability density function that a new $u_{ij}$ value corresponds to that ideological position. The green line represents the distribution of deputy $A$, with low standard deviation, which means that the votes of deputy $A$ are generally aligned with what the rest of the legislators of a similar ideological position vote. In contrast, the red line shows $B$ who, although ideologically seems to be close to $A$, has a much greater standard deviation. Therefore, deputy $B$ tends to vote in a much more fluctuating manner in the political spectrum, despite being ideologically similar to deputy $A$, that is, $d_{1B} \approx d_{1A}$.

Figure 1. Probability distribution of votes from two legislators

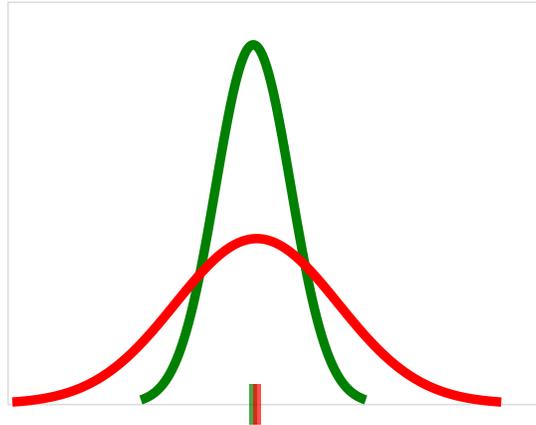

In green and red, the probability density function estimated from the votes of two Chilean legislators is graphed. While there are similarities in their ideological positions, defined here as the mean of the legislators' voting distribution, the distribution of the green legislator is much more stable, whereas the red legislator has more variable votes. This notion is captured by measuring the standard deviation of the distributions.

The previous case exemplifies the need to identify, across the political spectrum, which legislators have lower levels of cohesion. In this regard, it is proposed that the second dimension $d_{2i}$ associated with legislator $i$ be the standard deviation of the $u_{ij}$ values, as indicated by equation [eq:dim2].

$$d_{2i} = \sqrt{\frac{\sum_{j=1}^{n}(u_{ij} - d_{1i})^2}{n}}$$

### 3.3. The union of both dimensions: an example of B-Call

To understand the relationship between the two dimensions, figure [fig:tbcall] presents the interaction between the dimensions of ideological positioning and cohesion. This interaction reflects complex and nuanced dynamics of legislative behavior. By analyzing the first dimension, the ideological spectrum of the legislators is identified through the average of their votes, giving us a first look at their general political inclinations. In turn, the second dimension allows observing the cohesion of the legislators through the variability of their votes.

Figure 2. B-Call enables identification of different dynamics through the analysis of its two dimensions: ideological position y cohesion.

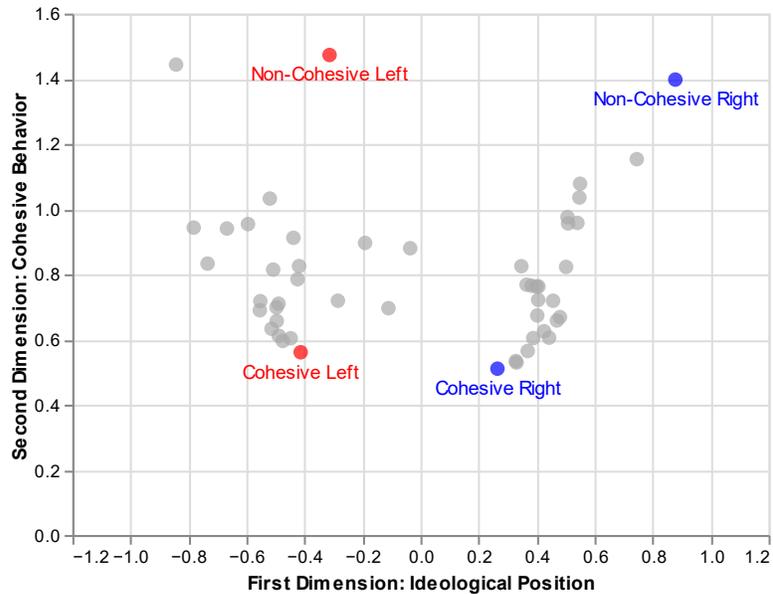

B-Call allows for the identification and quantification of different underlying behaviors in voting patterns. For instance, legislators with low standard deviation, depicted in our figure as red and blue dots labeled "Cohesive-Left" and "Cohesive-Right," demonstrate high cohesion in their voting patterns. This indicates that, despite their diverging political positions, both exhibit a significant level of predictability and uniformity in how they vote relative to their respective ideological stances .

In contrast, those legislators positioned at the top of the figure, showing high standard deviation, pose challenges to the traditional notion of ideological stance. These cases highlight the presence of legislators whose votes are ideologically marked but highly variable, suggesting an independence that could transcend known ideological divisions. This variability can be interpreted in various ways, including responses to specific pressures, increasing personalism, and fragmentation.

By integrating both dimensions, B-Call not only facilitates the identification of voting patterns but also provides a better understanding of legislative behavior in each congress. Quantifying voting variability, anchored to the ideological positioning, this variability, or lack thereof, offers a better understanding of the underlying political strategies at different levels, but mainly at the legislator level without the need for aggregation levels like parties or blocs.

## 4. Comparison with other models

This section will compare the two theoretically defined dimensions for B-Call with other widely used models for analyzing legislative behavior. For the ideological dimension (first dimension), we will examine the correlations of B-Call with the first dimension of W-

NOMINATE, while for the cohesion dimension (second dimension) we will contrast with UNITY and RICE indicators. For this comparison, we used voting information from the legislative chambers of three countries (United States, Brazil, and Chile) between 2003 and 2022.

*4.1. Correlation of B-Call with W-NOMINATE*

For each legislative chamber, the dimensions of our model and of W-NOMINATE were calculated. For this, we decided to analyze the data on an annual basis, which, in some cases in the Brazilian Senate, resulted in the inability of W-NOMINATE to calculate legislators' scores due to the low number of votes. For this model, legislators were divided into left and right according to the first dimension estimated by W-NOMINATE. This decision was made because the goal is to compare B-Call with W-NOMINATE estimations.

Firstly, the Pearson correlation and its standard errors between the first dimension of the model and the first dimension of W-NOMINATE were calculated. Overall, the results indicate a very strong positive correlation, suggesting they are significantly aligned in the scores associated with ideological position. On average, the highest correlations are found for the Chilean Chamber of Deputies and the U.S. House of Representatives. Brazil, comparatively, shows the worst average correlations.

Furthermore, the Spearman correlation between the first dimension of B-Call and the first dimension of W-NOMINATE was also calculated. The Spearman correlation corresponds to the Pearson correlation applied to the order in which each value appears, rather than the value itself. This allows us to analyze in more depth how both models compare when ideologically ranking legislators, regardless of the magnitude of each value. Several cases with low Pearson correlations are observed. All Spearman coefficients are very strong positives and exceed $\rho > 0.85$, indicating that both B-Call and W-NOMINATE produce a very similar ideological ordering.

The results of both correlations are found in the tables 1 and 2.

*Table 1. Pearson Correlation ($r$) and standard deviation (SE) between the first dimension of our model and W-NOMINATE*

|      | United States | | Brazil | | Chile | |
| --- | --- | --- | --- | --- | --- | --- |
| Year | $r$ | SE | $r$ | SE | $r$ | SE |
| 2022 | **0.964** | 0.013 | 0.986 | 0.007 | 0.991 | 0.011 |
| 2021 | 0.988 | 0.007 | 0.988 | 0.007 | 0.997 | 0.007 |
| 2020 | **0.938** | 0.017 | 0.984 | 0.008 | 0.997 | 0.006 |
| 2019 | 0.990 | 0.007 | 0.962 | 0.012 | 0.997 | 0.007 |
| 2018 | **0.972** | 0.011 | **0.880** | 0.021 | 0.993 | 0.010 |
| 2017 | 0.992 | 0.006 | 0.926 | 0.017 | **0.982** | 0.017 |
| 2016 | 0.995 | 0.005 | 0.961 | 0.012 | 0.982 | 0.018 |
| 2015 | 0.992 | 0.006 | 0.981 | 0.008 | **0.980** | 0.018 |
| 2014 | 0.988 | 0.007 | 0.944 | 0.017 | **0.980** | 0.018 |
| 2013 | 0.993 | 0.006 | 0.935 | 0.016 | **0.978** | 0.019 |

| Year | | | | | | |
|------|-------|-------|-------|-------|-------|-------|
| 2012 | 0.988 | 0.007 | **0.857** | 0.025 | 0.992 | 0.012 |
| 2011 | 0.995 | 0.005 | 0.908 | 0.018 | 0.987 | 0.015 |
| 2010 | **0.978** | 0.010 | **0.875** | 0.024 | 0.991 | 0.013 |
| 2009 | 0.991 | 0.007 | 0.921 | 0.017 | 0.991 | 0.013 |
| 2008 | 0.990 | 0.007 | 0.895 | 0.020 | 0.990 | 0.013 |
| 2007 | 0.997 | 0.004 | 0.947 | 0.014 | 0.993 | 0.011 |
| 2006 | 0.994 | 0.005 | 0.910 | 0.020 | 0.997 | 0.007 |
| 2005 | 0.996 | 0.004 | 0.900 | 0.021 | 0.987 | 0.015 |
| 2004 | 0.993 | 0.006 | **0.715** | 0.033 | 0.991 | 0.013 |
| 2003 | 0.996 | 0.004 | 0.885 | 0.042 | 0.996 | 0.008 |
| M    | 0.987 | 0.007 | 0.918 | 0.018 | 0.990 | 0.012 |
| SD   | 0.014 | 0.003 | 0.062 | 0.009 | 0.006 | 0.004 |

The four worst correlations ($r$) for each institution are shown in bold font.

*Table 2. Spearman correlation ($\rho$) and its standard deviation (SE) between the first dimension of our model and W-NOMINATE*

| | United States | | Brazil | | Chile | |
|------|-------|-------|-------|-------|-------|-------|
| Year | $\rho$ | SE | $\rho$ | SE | $\rho$ | SE |
| 2022 | **0.983** | 0.009 | 0.963 | 0.012 | 0.989 | 0.012 |
| 2021 | 0.985 | 0.008 | 0.966 | 0.011 | 0.989 | 0.012 |
| 2020 | **0.989** | 0.007 | 0.982 | 0.008 | 0.994 | 0.009 |
| 2019 | 0.987 | 0.008 | 0.962 | 0.012 | 0.981 | 0.016 |
| 2018 | **0.947** | 0.015 | **0.929** | 0.016 | 0.959 | 0.023 |
| 2017 | 0.952 | 0.015 | 0.948 | 0.014 | **0.960** | 0.026 |
| 2016 | 0.976 | 0.010 | 0.952 | 0.013 | 0.961 | 0.026 |
| 2015 | 0.984 | 0.009 | 0.972 | 0.010 | **0.961** | 0.026 |
| 2014 | 0.988 | 0.007 | 0.950 | 0.016 | **0.932** | 0.033 |
| 2013 | 0.992 | 0.006 | 0.936 | 0.016 | **0.968** | 0.023 |
| 2012 | 0.988 | 0.007 | **0.949** | 0.015 | 0.981 | 0.018 |
| 2011 | 0.992 | 0.006 | 0.920 | 0.017 | 0.970 | 0.022 |
| 2010 | **0.986** | 0.008 | **0.884** | 0.023 | 0.971 | 0.022 |
| 2009 | 0.994 | 0.005 | 0.890 | 0.020 | 0.958 | 0.026 |
| 2008 | 0.986 | 0.008 | 0.917 | 0.018 | 0.959 | 0.026 |
| 2007 | 0.996 | 0.004 | 0.893 | 0.020 | 0.943 | 0.031 |
| 2006 | 0.994 | 0.005 | 0.956 | 0.014 | 0.958 | 0.026 |
| 2005 | 0.991 | 0.006 | 0.945 | 0.016 | 0.940 | 0.032 |
| 2004 | 0.995 | 0.005 | **0.907** | 0.020 | 0.936 | 0.033 |
| 2003 | 0.989 | 0.007 | 0.845 | 0.048 | 0.951 | 0.029 |
| M    | 0.985 | 0.008 | 0.933 | 0.017 | 0.963 | 0.024 |
| SD   | 0.013 | 0.003 | 0.035 | 0.008 | 0.018 | 0.007 |

The same four worst years for each institution from Table 1 are shown in bold font.

*4.2. Cohesion dimension correlation (second dimension) with RICE and UNITY indicators*

This section addresses the correlation between the RICE and UNITY indicators and the cohesion dimension of B-Call. RICE and UNITY are traditional indicators that are based on a certain level of aggregation, usually at the party or coalition level. On the other hand, the cohesion dimension of the model proposed here has the advantage of providing an indicator at the individual level. This allows for capturing nuances and dynamics that are overlooked in aggregated analyses. For the three countries, we compared RICE and UNITY with the average of the B-Call cohesion indicator, according to the chosen level of aggregation. Annual voting levels were used.

*Table 3. Correlation of the second dimension of our model with RICE and UNITY in the United States, Brazil, and Chile*

|  | Party | RICE | | UNITY | |
|---|---|---|---|---|---|
|  |  | $r$ | SE | $r$ | SE |
| **USA** | Democrats | -0.848 | 0.125 | -0.725 | 0.162 |
|  | Republicans | -0.335 | 0.222 | -0.645 | 0.180 |
| **Chile** | Left | -0.601 | 0.188 | -0.286 | 0.226 |
|  | Right | -0.764 | 0.152 | -0.250 | 0.228 |
| **Brazil** | PTB | -0.620 | 0.185 | -0.195 | 0.231 |
|  | PSDB | -0.032 | 0.236 | -0.827 | 0.133 |

For the United States, indicators were calculated for both the Democratic and Republican parties. Table 3 shows a strong negative correlation with the cohesion dimension of B-Call for the Democratic party. The trend is similar for the Republican party, although slightly less in relation to RICE and significant with respect to UNITY

Measuring these traditional cohesion and unity indicators in Chile and Brazil presents additional complexities, mainly due to high levels of party fragmentation. This fragmentation highlights the limitations of aggregated indicators, as they depend on the level of aggregation established, typically party or coalition. Based on this difficulty, in the case of Chile, parties were classified into two broad ideological spectra (left and right). This classification allows tracing the temporal evolution. It is relevant to mention the composition of the legislative chamber, for example, in Chile, the number of represented parties increased significantly, from 7 (2002-2006) to 20 parties (2022). In Brazil, in 2019 alone, the composition was more than 30 parties.

In Chile, the data indicates a negative correlation with both conglomerates in RICE, being -0.764 for the right and -0.601 for the left. This is somewhat lesser with UNITY, mainly on the right (-0.250). For Brazil, there is more variability, for example, the nearly null correlation with RICE for the PSDB (-0.032) versus the high negative correlation with UNITY (-0.827).

The implications of these results can vary. Theoretically, this negative correlation is expected and logical, as higher RICE and UNITY values (indicating greater cohesion or unity within a group/party) are associated with a decrease in standard deviation. This is because a high degree of unity or cohesion implies that legislators vote more homogeneously following party or ideological lines. This suggests that legislators with low standard deviations tend to vote in accordance with their ideological positions and are probably more loyal to their party's directives. This supports the notion that the previously mentioned indicators successfully capture what they intend to, validating the cohesion dimension of B-Call in identifying legislative behavior patterns consistent with party loyalty and cohesion.

## 5. Empirical evaluation of B-Call and advantages over other models

*5.1. B-Call's contribution to calibrating the ideological positioning of legislators with low cohesion*

One of the strengths of B-Call is its ability to analyze the two main dimensions of legislative behavior and apply these analyses in both stable and fragmented decision-making environments. This advantage is significant. As noted by Krehbiel and Peskowitz , traditional indicators often encounter problems when parliamentary behaviors are less cohesive, inconsistent, or volatile. Thus, while a high correlation has been demonstrated between the first dimension of the indicator and the traditional W-NOMINATE, attention should also be directed to cases where this correlation is not direct, especially where legislators exhibit very low cohesion in their voting decisions.

Using data from the previously mentioned countries, figure 1 illustrates three cases where there is no correlation between our model and the W-NOMINATE indicator due to the low cohesion of their behavior. In this figure, the ten legislators with the lowest cohesion are highlighted in black. It's noteworthy that several legislators with the greatest discrepancies between the first dimension of B-Call and W-NOMINATE are those with lower group cohesion in voting. In simple terms, B-Call can help resolve the confusion of W-NOMINATE results for those congress members with high vote decision volatility.

Concrete examples of legislators whose votes did not follow a party cohesion line will be presented. These examples empirically demonstrate the strength of B-Call over traditional methods. The first is the analysis of the United States Congress in 2018. In this case, W-Nominate shows that U.S. Representative Justin Amash, a Republican and member of the Liberty Caucus until 2019, is positioned as a moderate within the Republican party despite being known as a far-right actor. Similarly, W-NOMINATE positioned two other legislators close to the Liberty Caucus, Thomas Massie and Andrew Bigg, nearer to the ideological center.

The voting analysis demonstrates that these representatives are less cohesive compared to their peers, distorting the positioning measured through traditional techniques. By using B-Call, the legislator's positioning is corrected for cohesion dimension, placing them more accurately ideologically. The same is observed for Brazil and Chile. In Brazil, despite a marked right-wing stance, W-NOMINATE placed Jair Bolsonaro in 2010 in a much more moderate ideological position, whereas our model considers him as one of the most conservative deputies with low cohesion levels. In Chile 2014, Gabriel Boric was seen by W-NOMINATE as a centrist within the left, while B-Call defines him to the left of the ideological spectrum. In both cases, a detailed inspection of their voting confirms B-Call as a viable alternative for measuring parliamentary behavior.

Figure 2. Discrepancies in the Ideological Estimation

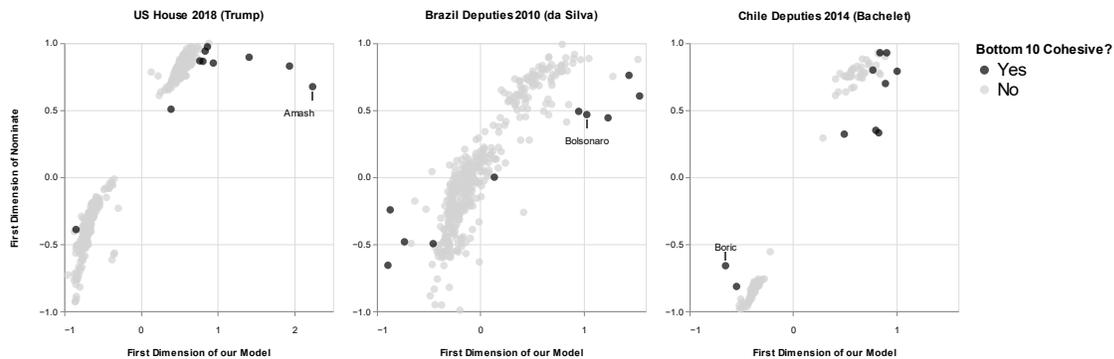

*5.2. Adaptability of B-Call to analyze different legislative contexts*

This section will evaluate the model in roll-call votes of the congresses of the United States, Brazil, and Chile over a 20-year period from 2003 to 2022, for both the lower and upper houses. This broad examination across different periods and legislative chambers in three countries aims to test the model's effectiveness in stable and fragmented congressional periods, demonstrating B-Call's empirical utility in capturing both the ideological positioning and the cohesion of legislators' voting.

The U.S. data were obtained from Lewis et al., (2023), while Brazilian and Chilean voting data were obtained from their respective official open data sections (Camara.leg.br, Senado.leg.br, Camara.cl, and Senado.cl). To avoid distortions in the analysis, legislators who voted in less than 10% of the cases were excluded. Moreover, the left-right division of legislators was determined using a clustering algorithm based on voting behavior (see appendix). In each institution, eight legislators were arbitrarily selected, either because of their relevance or their low cohesion.

*5.3. United States*

Analysis using B-Call across various legislative periods in the United States reveals dynamic changes in the House of Representatives' behavior throughout the first two decades of the 21st century. Figure 3 illustrates the division between Democrats and Republicans from 2003 to 2008 under the Bush administration, transitioning to a bridging of the ideological gap during the Obama years (2009-2014). However, in the last legislative period of Obama's presidency (2015-2016), the division between the two parties became pronounced again. Under the Trump administration, this division deepened with an increasing number of Republican representatives showing low cohesion. During the 2019-2020 legislative period, Democrats appeared organized and cohesive, while Republicans displayed greater ideological spread and lower cohesion, indicating emerging ideological dissonance within the Republican Party.

Figure 3. United States House of Representatives between 2003 and 2022

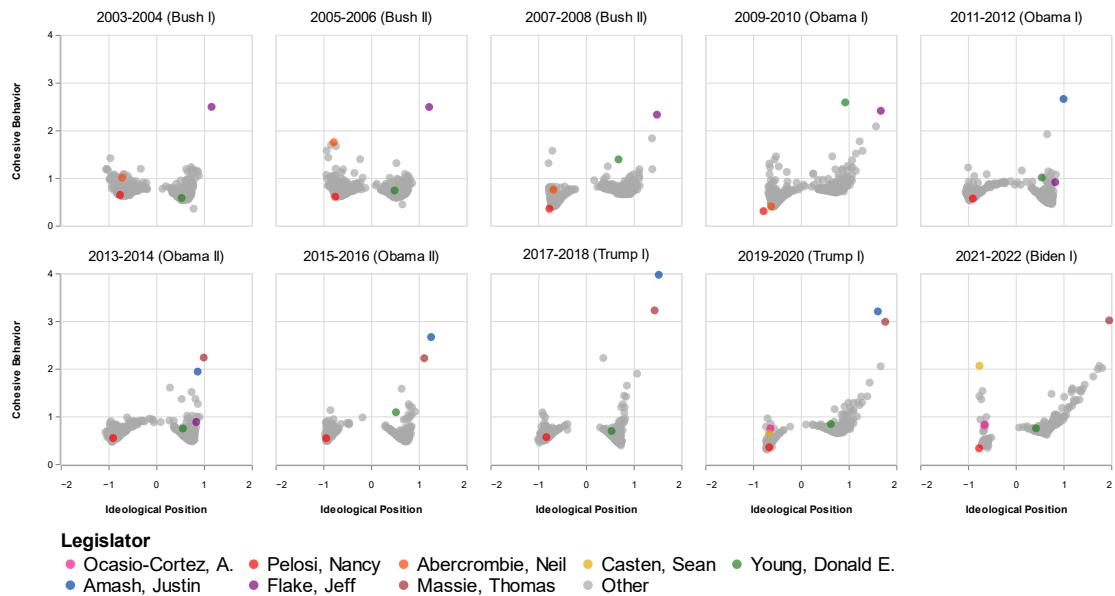

*Although Representative Rand Paul met the minimum number of votes between 2003 and 2010, he was excluded from these periods to improve the visualization. In these 4 legislative periods, he had low cohesion and was the most right-leaning legislator.*

Within the Democratic Party, Figure 4 highlights Nancy Pelosi and Alexandria Ocasio-Cortez. Over the 20 years analyzed, Pelosi has been in the House of Representatives, consistently aligning with the left wing of the Democratic Party and maintaining a high level of cohesion, meaning Pelosi has consistently voted in line with her faction. This contrasts with Ocasio-Cortez, who during the 2019-2020 and 2021-2022 periods has been among the less cohesive Democratic legislators. Notably, Representative Sean Casten was the Democrat with the lowest level of cohesion during 2021-2022, despite exhibiting similar voting behavior to Ocasio-Cortez in the previous period. Among Republicans, Representatives Justin Amash, Jeff Flake, Thomas Massie, and Donald Young stand out, as they tend to occupy the far-right wing of the party and exhibit low levels of cohesion.

B-Call was also tested on the evolution of the United States Senate. Figure 4 shows the results of a division between Democrats and Republicans during the Bush Administration. It also shows that in the last legislative period (2007-2008), a large part of the Democratic senators had cohesive behaviors, in contrast, the Republican senators were less so. This dynamic extends from 2007 to 2014. The analyses of the two-dimensional tool reflect that in the last legislative period (2015-2016) during the Obama Administration, the Senate was divided into similar poles between Democrats and Republicans. Then, during the Trump Administration, Republicans voted in a cohesive manner, while Democrats became more disordered. Finally, the legislative period 2021-2022 during the Biden Administration shows a relative symmetry between both parties. Both Democrats and Republicans present senators with low and high levels of cohesion, and there is a group of centrist legislators who bridge both political forces.

In Figure 4, eight senators are presented. Among the liberals, Bernie Sanders and Kamala Harris exhibit different voting behavior compared to Joe Biden. Sanders is not only positioned as one of the most liberal senators in the Senate but has also often been the least cohesive senator in the entire liberal sector. During her two terms, Harris similarly positioned herself on the far-left side of the Democratic Party, with low levels of cohesion. In contrast, Biden was positioned in the center of the Democratic Party between 2003 and 2006, though in his last period, 2007-2008, which coincided with his vice-presidential candidacy, he is located in the most liberal sector of the Democrats and with high cohesion. Among the Republicans, Senators Thomas Coburn, James DeMint, Mike Lee, and Rand Paul occupy the most extreme and least cohesive positions.

Figure 4. United States Senate between 2003 and 2022

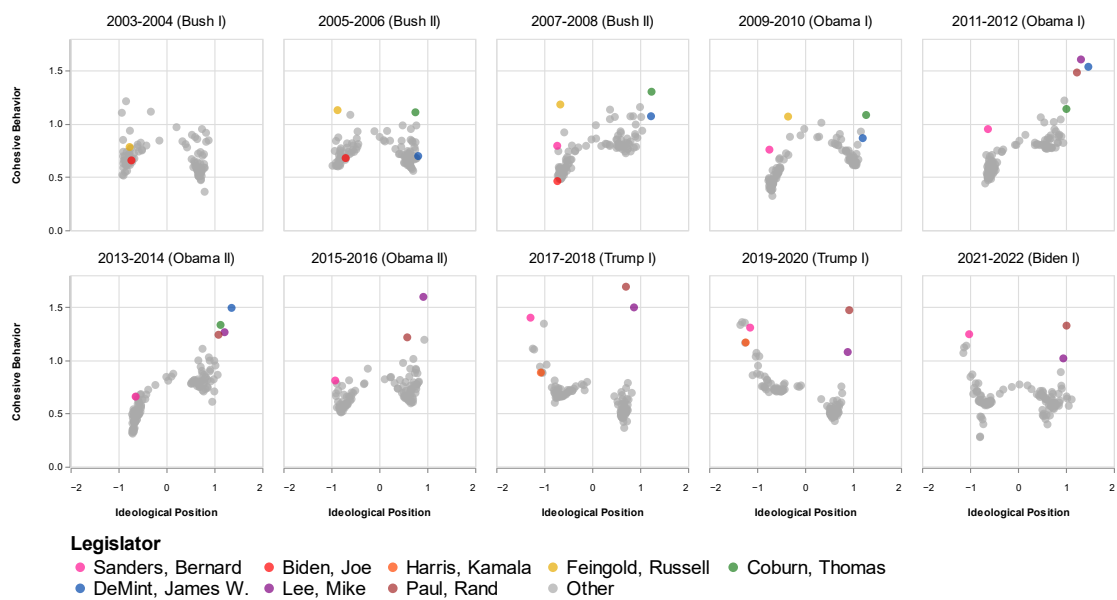

*The votes of Presidents Obama and Biden, as well as Vice President Harris, during their terms as executive authorities are not included.*

*5.4Brazil*

The Brazilian context differs completely from the U.S. one, especially due to the high fragmentation of its political system. In such a context, the strengths of B-Call are tested in a highly personalistic and volatile decision-making environment. This personalism is captured by the tool, as it notes diffuse divisions between political forces. Figure 4 shows Brazilian deputies grouped in a large cluster with low cohesion and identifies legislative periods, like 2007-2010 under Lula da Silva's administration and 2019-2022 under Bolsonaro's government, where divisions are observed. Similar to the deputies, the Brazilian Senate (Figure 5) does not show a clear separation between two political forces.

*Figure 5. Brazil Chamber of Deputies between 2003 and 2022*

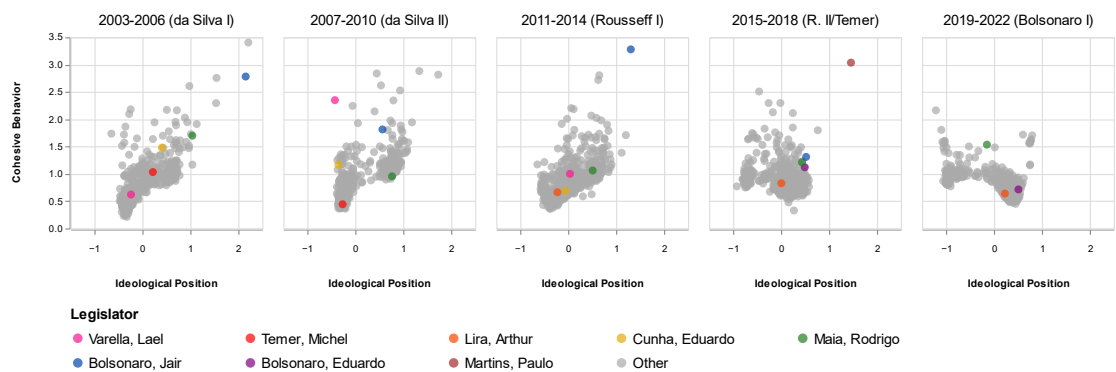

In individual analysis, Jair Bolsonaro presents as a deputy with very little cohesion toward his sector. During his tenure as a deputy between 2003 and 2018, he was mostly a parliamentary actor positioned further to the right of his sector and uncohesive. B-Call shows that Bolsonaro was one of the most right-leaning and least cohesive legislator according to his voting behavior.

*Figure 6. Brazil Senate between 2003 and 2022*

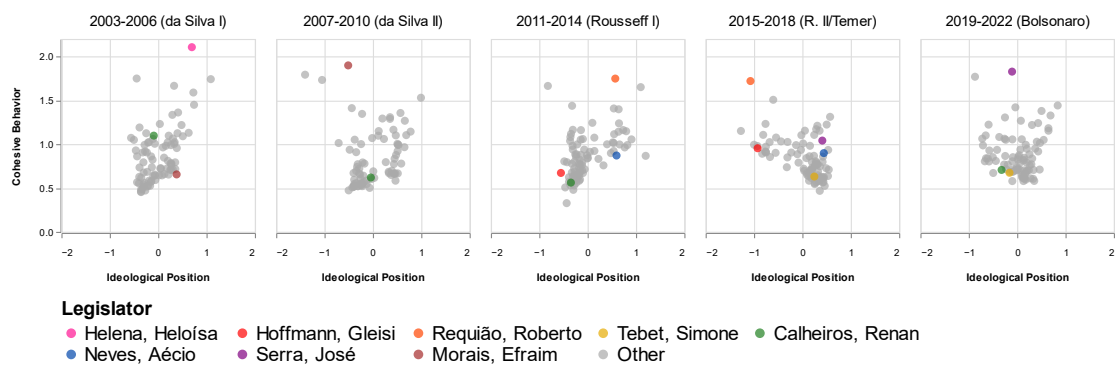

*Chile*

B-Call is then tested in Chile, a country that maintained relative stability within Congress until 2018 before exhibiting a high level of party fragmentation. The analysis, shown in Figure 7, reveals that under the Concertación governments (up to 2009), there was a clear division between the government (left) and opposition (right). It also shows that with the arrival of Piñera in 2010, the left experienced greater dispersion, with deputies positioned at the far-left extremes not previously seen in Congress. B-Call further indicates that cohesion problems began to surface before the electoral change. During Bachelet II, for instance, legislators with lower levels of cohesion appeared, becoming more evident in the last Piñera II government, where partisan divisions dissolved and personalism with low political cohesion began to emerge.

*Figure 7. Chile Chamber of Deputies between 2003 and 2022*

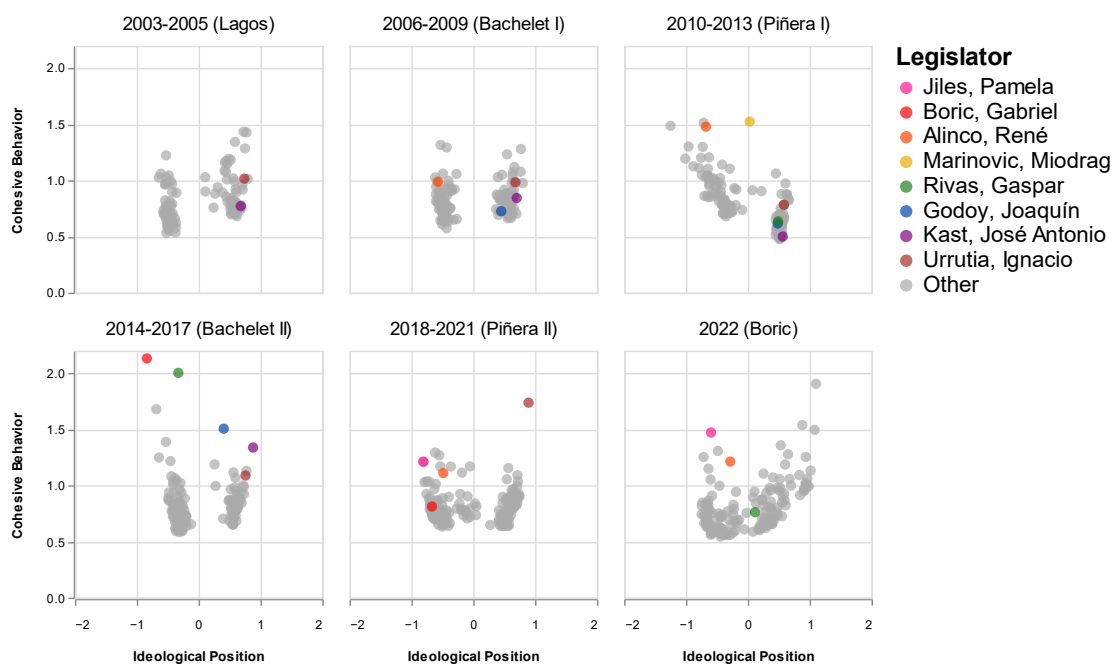

In the Chilean Senate, shown in Figure 7, there is notable disorder among senators with the arrival of the right with Piñera's Administration between 2010 and 2013. A group of senators in the center, a left sector with high dispersion in cohesion indicators, and a very cohesive right sector are observed. However, during Bachelet II, senators polarized into two factions, a pattern that recurs during Piñera II. In the first year of the Boric Administration, the left and right re-center, and the number of non-cohesive senators increases, especially on the right.

In terms of individual analysis, B-Call shows that during his first term (2014-2017), current President Gabriel Boric was the least cohesive deputy, positioned on the left of the ideological spectrum. In the following period, Deputy Boric demonstrated greater cohesion among his left-wing peers. In contrast, José Antonio Kast positioned himself among the deputies furthest to the right, albeit with fluctuating levels of cohesion. The 2014-2017 period is notable because Boric was the most left-leaning and least cohesive deputy, while Kast was the furthest to the right and the sixth least cohesive. In the Senate, Senators Adolfo Zaldívar and Lily Pérez stand out, as they occupied centrist positions during certain legislatures—under Bachelet I and Bachelet II, respectively—yet exhibited low levels of cohesion.

*Figure 8. Chile Senate between 2004 and 2022*

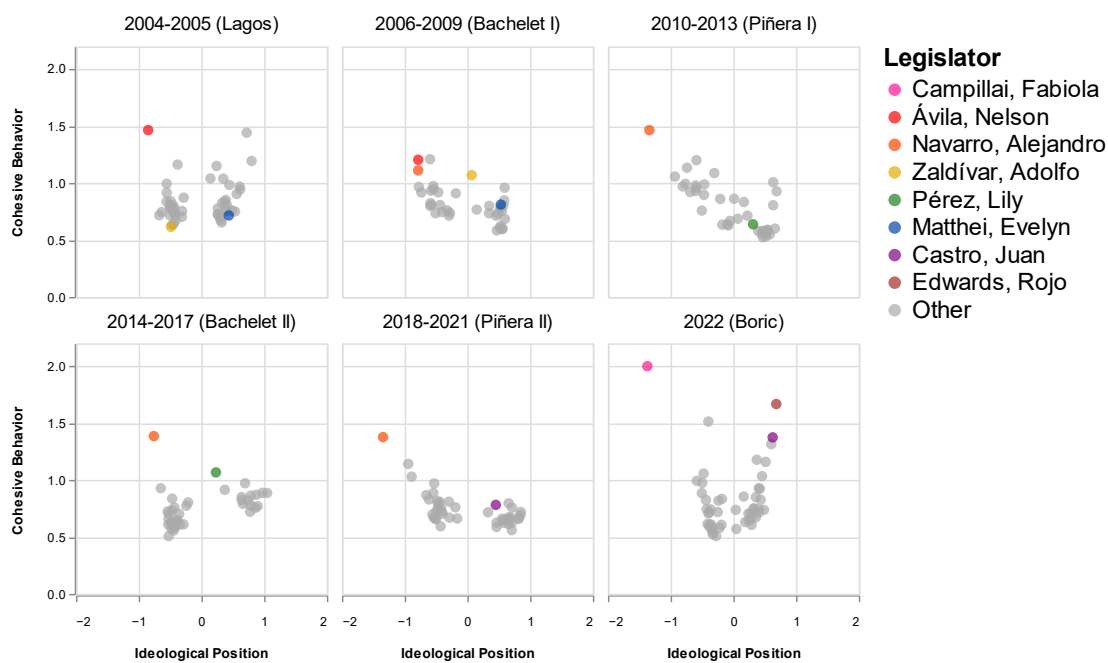

## Conclusion

B-Call continues a line of work in political science that seeks to estimate legislative behavior through data related to roll-call votes. B-Call possesses significant advantages over previous methods. This tool, for example, allows for the estimation of the ideological position alongside the level of cohesion that legislators exhibit in relation to that estimation.

Likewise, this paper demonstrated that it is possible to identify groups of actors with similar ideologies but with very different political behaviors, where some show high group cohesion, and others much less cohesive behavior. We also showed that B-Call can be used to estimate just one of the study dimensions, either ideology or political cohesion. Moreover, by comparing on one hand the ideological values of B-Call with the values of other important methods in the literature, and on the other hand the cohesion values with values from other methods used to measure unity, we see that B-Call delivers quite similar results in both dimensions.

Furthermore, particularly concerning the ideological dimension, we empirically demonstrated how some of B-Call's discrepancies with methods like W-NOMINATE illustrate how in certain cases it is possible to more accurately correct legislative behavior. Being a fully frequentist method, B-Call can also be used in cases with a poorer amount of data. Furthermore, B-Call can also be inside parties or parlamentary groups, wherein the first dimension represents the ideology of legislators within that group, and the second dimension represents the cohesion with respect to these parties.

Thus, B-Call provides a new tool for political analysis studies, especially when studying fragmented, volatile, or changing systems over time.

## Declaration of Conflicting Interests

The author(s) declared no potential interest with respect to the research, authorship, and/or publication of this article

## Funding

This work was supported by ANID – Millennium Science Initiative Program – Code ICN17-002 and NCS2022-046, and grant Fondecyt Nº 1211297.

## ORCID iDs

Juan Reutter https://orcid.org/0000-0002-2186-0312

Sergio Toro https://orcid.org/0000-0003-1129-5441

Lucas Valenzuela Everke

Daniel Alcatruz https://orcid.org/0000-0002-0323-1264

Macarena Valenzuela https://orcid.org/0000-0002-1707-5522

## Data Avaliability Statement

https://github.com/b-call-model/replication/tree/main

**Appendix 1. Dividing legislators into left and right**

According to Equation, the calculation of the values $u_{ij}$ requires knowing the votes of the left-wing and right-wing representatives in each vote $j$, to properly assign the sign of each $u_{ij}$. This division can be carried out manually based on the known political identification of the legislators or through unsupervised clustering algorithms such as Gaussian Mixture Model or Agglomerative Clustering .

Besides these models, an approach based on agglomerative clustering is proposed, which is suited to the nature of the problem of dividing legislators into two poles. For this purpose, the following premise is used: One is considered left if not right. Similarly, one is right if not left. Therefore, a legislator will be classified as left if their discrepancy with respect to the average of the left group is smaller than their discrepancy with respect to the average of the right. Following this logic, a pseudocode for the agglomerative clustering process is presented below, which starts with the two most distant legislators forming the basis of each of the two clusters, and then in each iteration adds a new legislator to either the left or right cluster.

Calculate matrix $M$ of distance between pairs of legislators $x, y$ according to the following formula, where $V$ is the voting matrix with $n$ votes.

$$dist(x,y) = \frac{\sum_{j=1}^{n}|V_{xj} - V_{yj}|}{n}$$

Select from M the pair of legislators $x', y'$ with greatest distance Incorporate legislator $x'$ into cluster $c_x$ and legislator $y'$ into $c_y$ Calculate centroids of clusters $c_x$ and $c_y$ Calculate matrix $\widetilde{M}$ of distance between each legislator without a cluster and each centroid Select from $\widetilde{M}$ the legislator with shortest distance to a certain cluster Add said legislator to the corresponding cluster and eliminate them from the legislators without cluster

The aforementioned procedure ensures an exhaustive allocation of all legislators to one of the two cluster groups, each corresponding to a different political affinity: left or right. Given that the clustering process is iterative, it is suggested to conduct a conclusive evaluation to corroborate that the legislators of a given cluster indeed exhibit a smaller discrepancy with the centroid of their group, as opposed to the other cluster.

Calculate centroids of clusters $c_x$ and $c_y$ Calculate distance of each legislator with $c_x$ and $c_y$ Identify legislators with shortest distance to the cluster that they do not belong to Change the cluster of the identified legislators Recalculate centroids of clusters $c_x$ and $c_y$ Recalculate distance of each legislator with $c_x$ and $c_y$ Identify legislators with shortest distance to the cluster they do not belong to

## Appendix 2. Second Dimension Correlation between B-Call and Nominate

Apart from the two Nominate dimensions, the algorithm also returns other additional indicators used in the simulation that may provide valuable insights about the legislators. We are interested in determining whether any of these have a strong correlation with our second dimension. Table 4 presents the 10 indicators that we will analyze against our second dimension, which are explained in greater detail in the algorithm's documentation.

*Nominate indicators*

| | |
|---|---|
| coord1D | First dimension W-NOMINATE |
| coord2D | Second dimension W-NOMINATE |
| se1D | Bootstrapped standard error of first dimension W-NOMINATE |
| se2D | Bootstrapped standard error of second dimension W-NOMINATE |
| GMP | Geometric Mean Probability |
| PRE | Proportional Reduction In Error |
| correctYea | Predicted Yeas and Actual Yeas |
| wrongYea | Predicted Yeas and Actual Nays |
| wrongNay | Predicted Nays and Actual Yeas |
| correctNay | Predicted Nays and Actual Nays |

We chose to work with the 20 congressional sessions of the Chilean Chamber of Deputies. Procedurally, each simulation had to be run with the input parameter trials set to 6 for the algorithm to return the se1D and se2D indicators. All other input parameters were kept at their default values.

Table 6 shows the correlations for each legislature between B-Call's second dimension and each of the 10 Nominate indicators. Additionally, Table 5 summarizes these results using various statistics. As observed, all the indicators exhibit a high standard deviation, indicating a significant variability in the correlations. For example, while wrongYea has an average positive correlation of 0.518, there is considerable volatility in the individual correlations. In 2013, wrongYea had a correlation of 0.841 with B-Call's second dimension, but in 2006, it was only 0.194. The high standard deviation of all the indicators suggests that none of them consistently maintain a strong correlation with B-Call's second dimension. All show high volatility, as represented by the standard deviation, indicating that B-Call's second dimension is a novel indicator not captured by any Nominate indicator.

*Statistics of the correlations*

| Stats | cYea | wYea | wNay | cNay | GMP | CC | coord1D | coord2D | se1D | se2D |
|---|---|---|---|---|---|---|---|---|---|---|
| mean | -0.366 | 0.518 | 0.269 | 0.239 | -0.581 | -0.530 | -0.094 | -0.023 | -0.096 | -0.115 |
| std | 0.192 | 0.181 | 0.204 | 0.256 | 0.156 | 0.171 | 0.404 | 0.216 | 0.287 | 0.135 |
| min | -0.693 | 0.194 | -0.082 | -0.231 | -0.859 | -0.791 | -0.683 | -0.389 | -0.638 | -0.301 |
| 25% | -0.508 | 0.372 | 0.111 | 0.068 | -0.665 | -0.637 | -0.435 | -0.152 | -0.221 | -0.206 |
| 50% | -0.349 | 0.571 | 0.260 | 0.193 | -0.571 | -0.530 | -0.085 | -0.034 | -0.086 | -0.128 |
| 75% | -0.269 | 0.607 | 0.439 | 0.419 | -0.455 | -0.413 | 0.253 | 0.136 | 0.122 | -0.023 |
| max | 0.041 | 0.841 | 0.665 | 0.691 | -0.292 | -0.269 | 0.483 | 0.444 | 0.447 | 0.118 |

*Correlations of the second dimension of B-Call with Nominate indicators*

| Year | cYea | wYea | wNay | cNay | GMP | CC | coord1D | coord2D | se1D | se2D |
|---|---|---|---|---|---|---|---|---|---|---|
| 2003 | -0.278 | 0.570 | 0.270 | 0.047 | -0.635 | -0.576 | 0.116 | -0.286 | -0.130 | -0.214 |
| 2004 | -0.506 | 0.368 | 0.250 | 0.390 | -0.671 | -0.521 | 0.416 | -0.119 | 0.123 | -0.143 |
| 2005 | -0.399 | 0.373 | 0.188 | 0.192 | -0.600 | -0.426 | 0.410 | -0.207 | -0.023 | -0.301 |
| 2006 | -0.340 | 0.194 | 0.043 | 0.074 | -0.492 | -0.269 | -0.396 | 0.158 | 0.070 | 0.083 |
| 2007 | -0.236 | 0.449 | 0.052 | 0.109 | -0.439 | -0.383 | -0.124 | 0.145 | -0.164 | -0.136 |
| 2008 | 0.041 | 0.572 | 0.274 | -0.231 | -0.542 | -0.558 | 0.234 | 0.280 | 0.223 | -0.023 |
| 2009 | -0.272 | 0.244 | -0.082 | 0.142 | -0.419 | -0.271 | -0.195 | 0.082 | -0.009 | 0.118 |
| 2010 | -0.514 | 0.628 | 0.456 | 0.507 | -0.661 | -0.629 | -0.553 | -0.022 | -0.540 | -0.119 |
| 2011 | -0.580 | 0.774 | 0.558 | 0.564 | -0.792 | -0.784 | -0.651 | -0.064 | 0.447 | -0.287 |
| 2012 | -0.621 | 0.777 | 0.665 | 0.553 | -0.778 | -0.791 | -0.649 | -0.038 | -0.638 | -0.296 |
| 2013 | -0.637 | 0.841 | 0.543 | 0.668 | -0.825 | -0.786 | -0.683 | -0.134 | -0.598 | -0.295 |
| 2014 | -0.693 | 0.596 | 0.434 | 0.691 | -0.859 | -0.747 | -0.621 | -0.359 | 0.180 | -0.104 |
| 2015 | -0.460 | 0.343 | 0.471 | 0.382 | -0.539 | -0.540 | -0.046 | 0.444 | -0.159 | -0.075 |
| 2016 | -0.035 | 0.594 | 0.403 | -0.067 | -0.663 | -0.630 | 0.311 | -0.263 | -0.323 | -0.200 |
| 2017 | -0.220 | 0.605 | 0.286 | 0.011 | -0.617 | -0.660 | 0.065 | -0.389 | -0.294 | -0.082 |
| 2018 | -0.261 | 0.535 | 0.037 | 0.194 | -0.405 | -0.397 | -0.294 | 0.132 | 0.205 | 0.103 |
| 2019 | -0.273 | 0.611 | 0.099 | 0.143 | -0.487 | -0.466 | -0.264 | -0.055 | -0.042 | -0.154 |
| 2020 | -0.359 | 0.584 | 0.154 | -0.013 | -0.456 | -0.485 | 0.207 | 0.089 | -0.197 | 0.051 |
| 2021 | -0.397 | 0.439 | 0.115 | 0.204 | -0.452 | -0.418 | 0.350 | -0.031 | 0.122 | -0.204 |
| 2022 | -0.281 | 0.256 | 0.168 | 0.217 | -0.292 | -0.273 | 0.483 | 0.169 | -0.172 | -0.024 |